\documentclass[12pt,preprint]{aastex}

\shorttitle{Spiral Density Waves in M81}
\shortauthors{Wang et al.}
\usepackage{bm}
\usepackage{sansmath}
\usepackage{mathrsfs}
\usepackage{epstopdf}
\usepackage[usenames,dvipsnames]{color}
\usepackage[breaklinks,colorlinks,urlcolor=blue,citecolor=blue,linkcolor=blue]{hyperref}
\newcommand{\ron}[1]{{\color{black} #1}}
\newcommand{\hhwang}[1]{{\color{black} #1}}
\begin{document}

\title{Spiral Density Waves in M81. I. Stellar Spiral Density Waves}

\author{Chien-Chang Feng\altaffilmark{1}, Lien-Hsuan Lin\altaffilmark{1}, Hsiang-Hsu Wang\altaffilmark{1} and Ronald E. Taam\altaffilmark{1,2}}
\email{hhwang@asiaa.sinica.edu.tw}
\altaffiltext{1}{Institute of Astronomy and Astrophysics, Academia Sinica, P.O. Box 23-141, Taipei 10617, Taiwan, R.O.C.}
\altaffiltext{2}{Department of Physics and Astronomy, Northwestern University, 2131 Tech Drive, Evanston, IL 60208, USA}

\begin{abstract}
Aside from the grand-design stellar spirals appearing in the disk of M81, a pair of stellar spiral arms 
situated well inside the bright bulge of M81 has been recently discovered by \citet{Ken08}. The 
seemingly unrelated pairs of spirals pose a challenge to the theory of spiral density \ron{waves}. To 
address this problem, we have constructed a three component model for M81, including the contributions 
from a stellar disk, a bulge, and a dark matter halo subject to observational constraints. Given 
this basic state for M81, a modal approach is applied to search for the discrete unstable spiral modes 
that may provide an understanding for the existence of both spiral arms. It is found that the apparently 
separated inner and outer spirals can be interpreted as a single trailing spiral mode. In particular, 
these spirals share the same pattern speed 25.5~km~s$^{-1}$~kpc$^{-1}$ with a corotation radius of 9.03~kpc. 
In addition to the good agreement between the calculated and the observed spiral pattern, the variation of 
the spiral amplitude can also be naturally reproduced.
\end{abstract}


\section{Introduction}

The magnificent spiral arms observed in disk galaxies have been proposed to be a quasi-stationary phenomenon 
of a wave since the mid-1960s \citep{Lin64,Lin66}. The wave interpretation of spirals resolves the so-called 
winding dilemma problem that spirals would otherwise rapidly wrap up if a spiral is made of the same material 
in a differentially rotating disk. The stellar spiral arms are assumed to be a small perturbation superimposed 
on an basic state and, therefore, can be understood with linear analysis. Within the framework of a density wave, 
the shape of spirals is determined by a dispersion relation, which is an intrinsic property of a self-gravitating 
disk. Soon in the late 1960s, the working hypothesis of quasi-stationary spiral structure (QSSS) led to the study 
of the non-linear gas response to the underlying stellar spirals. The bright new born stars along spirals are 
interpreted to be a result of gas compression as molecular clouds pass through galactic shocks \citep{Rob69}. 
However, it was immediately recognized that the picture of the spiral density wave was not complete \citep{Too69}. 
The spiral structures are bound to disappear quickly as the wave packets propagate away from the disk. Hence, 
a feedback mechanism would be required to support the QSSS hypothesis. 

In 1970s, both in hydrodynamics and stellar dynamics, a self-consistent theory of global spiral mode was developed 
to explain the excitation and maintenance of density waves. Unlike the early development of spiral density wave 
theory in which the pattern speed is determined empirically or inferred from observations, the modal approach 
associates the growth rates and the pattern speeds to the discrete eigenvalues that are subject to boundary 
conditions. The spiral patterns are identified as the corresponding eigen-functions of modes. In other words, 
spiral arms are considered to be the self-excited unstable modes with calculated pattern speeds and growth 
rates. The mechanisms of a wave amplification loop have been analytically explored in great detail by several 
authors \citep{Mar76b,Mar77,Lin79,Ber89,Ber96,Ber00}. 

\hhwang{Given various types of spiral galaxies, the above mentioned classical density wave theory, which 
views the grand design spirals as intrinsic unstable modes of disks, is by no means the only possibility 
for the self excitation of spirals. Spirals can be interpreted as collective responses to clumps in 
a density distribution \citep{Gol65,Jul66,Too91} and understood in terms of the so-called 
swing-amplification mechanism \citep{Too81}. For example, the global $N$-body simulations conducted by 
\citet{Sel12} show that the growth of non-axisymmetric waves can be characterized by two phases. The first, 
slowly growing phase is due to the swing-amplified shot noise that propagates inwards and is absorbed at 
the inner Lindblad resonance. The accumulating change in the distribution function at this resonance, owing 
to wave-particle interactions, leads to the second phase corresponding to the rapid growth of 
non-axisymmetric spirals. The exposure of the inner Lindblad resonance to incident wave trains is crucial 
for the operation of this mechanism. More recently, \citet{Don13} demonstrate that apparent long-lived 
spirals can be a result of the non-linear development of gravitational wakes, which are initially induced 
by density clumps, such as giant molecular clouds. Once these wakes are formed, they may act as new 
perturbers to maintain the spiral forming activities. In this view, the grand design spirals are nothing 
more than connections between self-perpetuating wave segments so that the visually `long-lived' structures 
are understood in a statistical sense. This mechanism may potentially play a role in forming spiral 
structures recently observed in red disk galaxies with no gas \citep{Mas10} and in dwarf elliptical 
galaxies in the Virgo cluster \citep{Lis06, Lis09}. Here, we do not attempt to complete the list of 
possible mechanisms as a more thorough discussion on this topic has been recently reviewed by \citet{Sel13}. 
However, we believe that different theories or mechanisms may operate in different types of spiral galaxies 
during secular evolution. Distinguishing between different possibilities for a specific galaxy can only 
rely on qualitative and quantitative comparisons between theoretical predictions and observations. In the 
following, we apply the modal analysis to the galaxy M81 within the framework of spiral density wave 
theory to confront the observations.}

M81 (NGC 3031) is one of the most well-studied spiral galaxies due to its proximity to the Milky Way and 
its good inclination, which is suitable for extracting kinematic information. The presence of a central 
bright bulge together with the nearly bi-symmetric spiral arms found in the galactic disk make M81 one 
of the best targets for investigating the origin of spiral structure. \citet{Vis80b} applied the dispersion 
relation to a basic model for M81 to fit the shape of stellar density wave observed outside the radius of 
3.5 kpc. The pattern speed, 
$\Omega_p=18$ km s$^{-1}$ kpc$^{-1}$, was chosen based on the morphology of spirals and also artificially 
constrained by the computational consideration for the non-linear gas response discussed in the second paper 
\citep{Vis80a}. In this particular model, density wave theory failed to explain structures situated inside 
the inner Lindblad resonance (ILR), which is located at 2.5 kpc. 
 
\citet{Low94} revisited the problem of spiral density waves in M81 based on the modal approach. A set of 
formulae was proposed to fit the observed rotation curve and the mass distribution of the disk in the range 
of $4<r<15$ kpc. The authors interpreted the variation of the spiral amplitude observed in the {\it I}-band 
image \citep{Elm89} as a result of the interference of trailing and leading waves that are propagating in the 
opposite directions. In their best-fit model, the pattern speed is determined to be $\Omega_p=24$ km s$^{-1}$ 
kpc$^{-1}$ with the corresponding corotation radius located at 9 kpc. The pattern speed is in good agreement with 
value, $\Omega_p=26$ km s$^{-1}$ kpc$^{-1}$, determined by the  non-linear response of gas at the 4:1 
resonance \citep{Elm89} as well as with the value, $\Omega_p=23.4 \pm 2.3$ km s$^{-1}$ kpc$^{-1}$, obtained by 
integration of the continuity equation \citep{Tre84,Wes98}.  However, the presence of the inner turning point 
at 4 kpc again rules out the possibility of wave propagation in the central region $r<4$ kpc. As also cautioned 
in \citet{Low94}, their analysis might be biased by the lack of reliable data in the red and infrared 
wavelengths at that time. 
 
Recently, \citet{Ken08} used {\it Spitzer} IRAC 3.6 and 4.5 $\mu$m near-infrared data from the Spitzer Infrared 
Nearby Galaxies Survey(SINGS), optical $B$, $V$ and $I$ and Two-Micron All-Sky Survey $K_s$-band data to trace 
the spiral density waves in M81. Using the data of different wavelengths enabled the authors to produce 
maps of stellar mass surface density and remove possible contaminations arising from young stars and from 
polycyclic aromatic hydrocarbons (PAHs). The non-axisymmetric structures (residuals) are then extracted by 
subtracting the axisymmetric component from the original maps. As the residual map of IRAC 3.6~$\mu$m shown 
in Fig.~\ref{fig:M81_residual_map}, aside from the outer stellar spirals, which are usually the primary 
focus in previous studies of M81, another pair of stellar spirals is revealed well inside the bright bulge of 
M81. As already noted in \citet{Ken08}, both models as discussed in \citet{Vis80b} and \citet{Low94} fail to 
explain the presence of the inner stellar spirals. In these theoretical investigations, the bulge region 
is considered dynamically hot and is, therefore, hostile to the presence of density waves. 

Furthermore, as also shown in \citet{Ken08}, a good correlation is found between the inner dust spirals (IRAC 
8~$\mu$m) and the inner stellar spirals (IRAC 3.6~$\mu$m). Since within the framework of density wave theory
where density waves in the gas are allowed to propagate through the inner Lindblad resonance, \citet{Ken08} 
interpreted the stellar spirals as a result induced by the gas spirals. However, from the dynamical point of 
view, it is unlikely that the inner stellar spirals can be induced by the gas spirals for the following reasons. 
First, the velocity dispersion in stars is much larger than that of gas in the bulge region. If we adopt 
the values from \citet{Low94}, the radial velocity dispersion in stars can be easily more than 40 km s$^{-1}$ 
(cf. for the central region of the Milky Way) in the central region in contrast to the value $\approx 10$ km 
s$^{-1}$ for the gas. Since the energy involved is proportional to the square of velocity dispersion, 
transferring energy from the gas to stars is not effective in perturbing the stellar system.  Second, little 
amount of gas is found in the bulge compared to the stellar mass. The total mass of the widespread HI in M81 is 
estimated to be $2.6\times 10^9$ ${\rm M}_{\odot}$ \citep{Yun94} in contrast to the more concentrated 
bulge mass enclosed within 3~kpc of $1\times 10^{10}$ ${\rm M}_{\odot}$ \citep{Low94}. Third, the 
thickness of the gas disk is much thinner than that of the stellar disk. In other words, the density waves in 
gas only effectively influence stars that are close to the mid-plane. Fourth, the absence of an ILR as shown 
in \citet{Wes96} does not support the possibility of wave excitation in gas at the radius of resonance. The 
non-linear gas response is basically forced by the underlying stellar spirals and belongs to the type of waves 
discussed in \citet{Rob69} (see also \citet{Fuj66,Shu73,Woo75}).  

The models proposed by \citet{Low94} and \citet{Vis80b} are not compatible to the presence of the newly 
discovered inner spiral structure.  With little details regarding the inner stellar spirals given in the 
literature, the origin and the property of the inner stellar spirals is elusive. Despite these difficulties, 
the faint tails of the inner spirals as shown in Fig.~\ref{fig:M81_residual_map} hint that the inner and outer 
spirals may be a single spiral mode. The seemingly separate structure is simply due to low density contrast if 
one compares the amplitude of the spirals to the bright axi-symmetric background. If this is a plausible 
interpretation, the following can be inferred. First, the inner and outer arms share the same pattern speed, 
which has been more or less constrained in the literature for the outer spirals. Second, from a dynamical 
consideration, the ILR may be absent as discussed in \citet{Adl96}, and the turning point has to be placed 
roughly at $r=1.4$~kpc. The location of the turning point as so chosen in \citet{Low94} is somewhat driven by 
the requirement of wave behavior, i.e., the wave is not allowed to propagate in the region of $r<4$~kpc. In 
general, the presence of a turning point is not guaranteed for a given rotation curve. Third, the coincidence 
of gas(dust) and stellar spirals is a natural outcome of the theory of spiral density waves. The non-linear 
gas response to the underlying stellar potential is an active subject in the field. 

Unlike the work of \citet{Low94}, who use heuristic formulae to fit the rotation curve of M81 for the 
range $r>4$~kpc, we start with a three-component model associated with a bulge, a dark halo, and an active 
stellar disk. The mass model is constructed to fit the whole range of the observed rotation curve and to fit 
the observed mass distribution. The resultant rotation curve is then treated as input for the modal analysis.
With this procedure and proper boundary conditions, we obtain unstable spiral modes that may explain the 
existence of spirals observed in M81. This paper is structured as follows. The three component model and 
the procedure of modal analysis are detailed in \S 2. In \S 3, we present the results obtained from the modal 
analysis and its comparison with observational data.  The results are discussed in \S 4 and we summarize in the last section.

\section{Method of Analysis}
\subsection{Observational Constraints and A Three-component Model for M81}

We apply the modal approach described in \citet{Lau78} to a zero-thickness fluid disk model for M81 to search 
for discrete global unstable modes that may account for the seemingly separated inner and outer spirals. 
The prominent bright bulge of M81 and the presence of inner spirals suggest that the central region is probably 
composed of a dynamically hot bulge and a relatively cold disk, within which the density wave can propagate. It 
is also well established that the rotation in the outer region of a disk galaxy is usually dynamically supported 
by a dark halo. Thus, the basic state of M81 is taken to be composed of an active stellar disk and two 
inactive spherical 
components, a bulge and a dark matter halo, which are usually dynamically hot and not involved in the 
propagation of spiral waves. The well-observed rotation curve is therefore associated with a three-component 
model composed of a stellar disk, a bulge, and a dark halo. 
 
The formula used to describe the density distribution of the spherical halo is given by \citep{Low94}:
\begin{equation} \label{equation:halo}
  \rho_{\rm H}(R) = \frac{\rho_{{\rm H}0}}{1+(R/R_{\rm H})^2},
\end{equation}
with $\rho_{{H}0}=3.5\times 10^7$ ${\rm M}_{\odot}$~kpc$^{-3}$ corresponding to the central volume density, 
$R$ the distance and $R_{\rm H}=2.8$~kpc a length scale controlling the concentration of the halo. The total mass 
$M_{\rm H}$ enclosed within the radius of 12 kpc is $2.84\times 10^{10}{\rm M}_{\odot}$. For the model of the 
bulge, we adopt the form described in \citet{Ath92},
\begin{equation} \label{equation:bulge}
\rho_{\rm b}(R) = \rho_{{\rm b}0}\left(1+\frac{R^2}{R_{\rm b}^2} \right)^{-1.5}
\end{equation}
with $\rho_{{\rm b}0}=5.5\times 10^8$ ${\rm M}_{\odot}$~kpc$^{-3}$ corresponding to the central volume 
density and $R_{\rm b}=1$~kpc, the concentration factor of the bulge. The total mass of the bulge enclosed 
within the radius of 12 kpc is $M_{\rm b}=1.5\times 10^{10}{\rm M}_{\odot}$. As for the surface density of the 
active stellar disk, we adopt the combination of two Toomre's disks \citep{Too63, Lau78},
\begin{equation} \label{equation:disk}
  \sigma_0(r) = \frac{4.5M_1}{\pi a_1^2}\frac{1}{x^{11}_1} - \frac{4.5M_2}{\pi a_2^2}\frac{1}{x^{11}_2}
\end{equation}
with $M_1=6.9\times 10^{10}$ ${\rm M}_{\odot}$, $M_2=7\times 10^9$ ${\rm M}_{\odot}$, $a_1=10$ kpc, $a_2=7$ kpc, 
$x_j=[1+(r/a_j)^2]^{1/2}$, $j=1,2$, and $r$ the radial distance measured in the disk plane. The mass in the 
active stellar disk, $M_{\rm d}$, is $6.2\times 10^{10}{\rm M}_{\odot}$. The scale length of the active disk 
fitted to an exponential profile for the range between 3~kpc and 10~kpc is estimated to be 2.5~kpc. This 
value is in excellent agreement with the observational value of 2.5~kpc \citep{Ken87} and is consistent 
with the value 2.83~kpc adopted in \citet{Low94}. Comparisons of our mass model with other works are summarized in 
Table~\ref{table:mass}. 
 
\begin{figure}
\epsscale{1}
\plotone{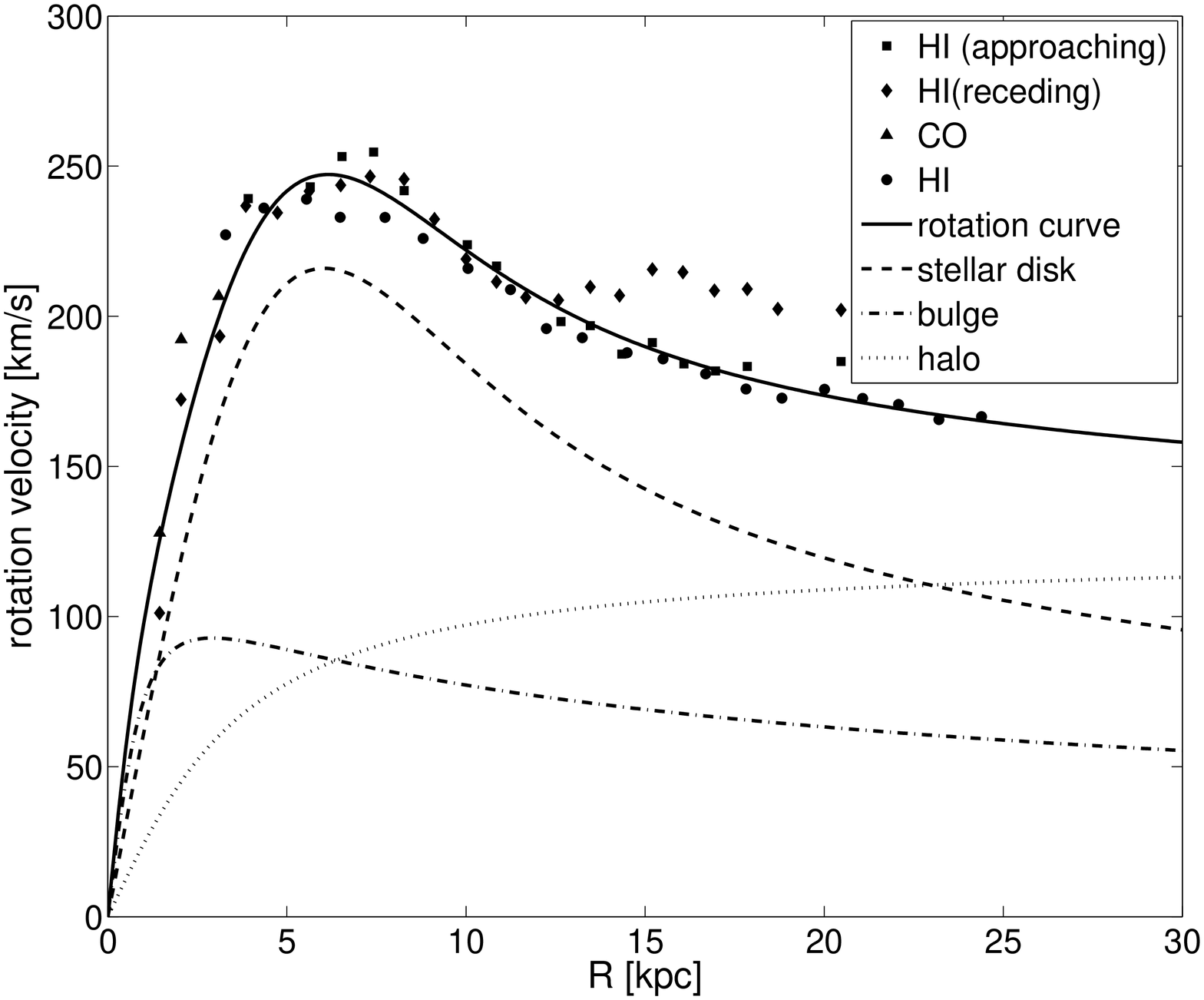}
\caption{The rotation curve of M81. The filled circles are the HI data tabulated in \citet{Ken87}. The diamonds 
and squares represent the HI observation on the receding and approaching sides, respectively \citep{Adl96}. The 
CO data for the inner part of M81 are the triangles \citep{Sag91}. The resultant rotation curve (solid line) 
resulting from contributions due to a stellar disk (dashed line), a bulge (dash-dotted line) and a dark matter 
halo (dotted line) is shown. The mathematical formulae used for these models are described in Sec.~2.1. 
\label{fig:rotation_curve}}
\end{figure}

Figure~\ref{fig:rotation_curve} shows the observed rotation curve as well as the fitting curve using the 
three-component model and the above parameters. The observed rotational velocity in the outer disk is 
traced using neutral hydrogen \citep{Adl96,Ken87}, while for the inner disk a few data points from the CO 
observation are adopted \citep{Sag91}. In this plot, all the observational data points have been normalized 
for the distance of M81 adopted in \citet{Ken08}, i.e., 3.6~Mpc. The contributions from the disk, the bulge 
and the dark halo are presented in dashed, dash-dotted and dotted lines, respectively. The fitting rotation 
curve, shown as the solid line, is in good agreement with the observational data. The reason of choosing 
these models is to ensure the central part of the M81 behaves as a solid body rotating with nearly the same 
angular speed. This feature is important for circumventing the singularity that would otherwise 
occur in the modal analysis (details described below). \hhwang{We avoid the use of popular models for the shape of the dark matter halo such as NFW \citep{Nav97} or Hernquist \citep{Her90} profiles for the following reasons. First, both of them have a density cusp at the center of galaxies which is not directly supported by observations. On the contrary, a constant dark matter density seems to be preferred in the inner parts of galaxies \citep{deB10} and is, therefore, well described by Eq.~(\ref{equation:halo}). Secondly, and more 
important, is that both profiles leads to infinite angular speed as the galactic center is approached, 
which is physically unrealistic.}

\begin{figure}
\epsscale{1}
\plotone{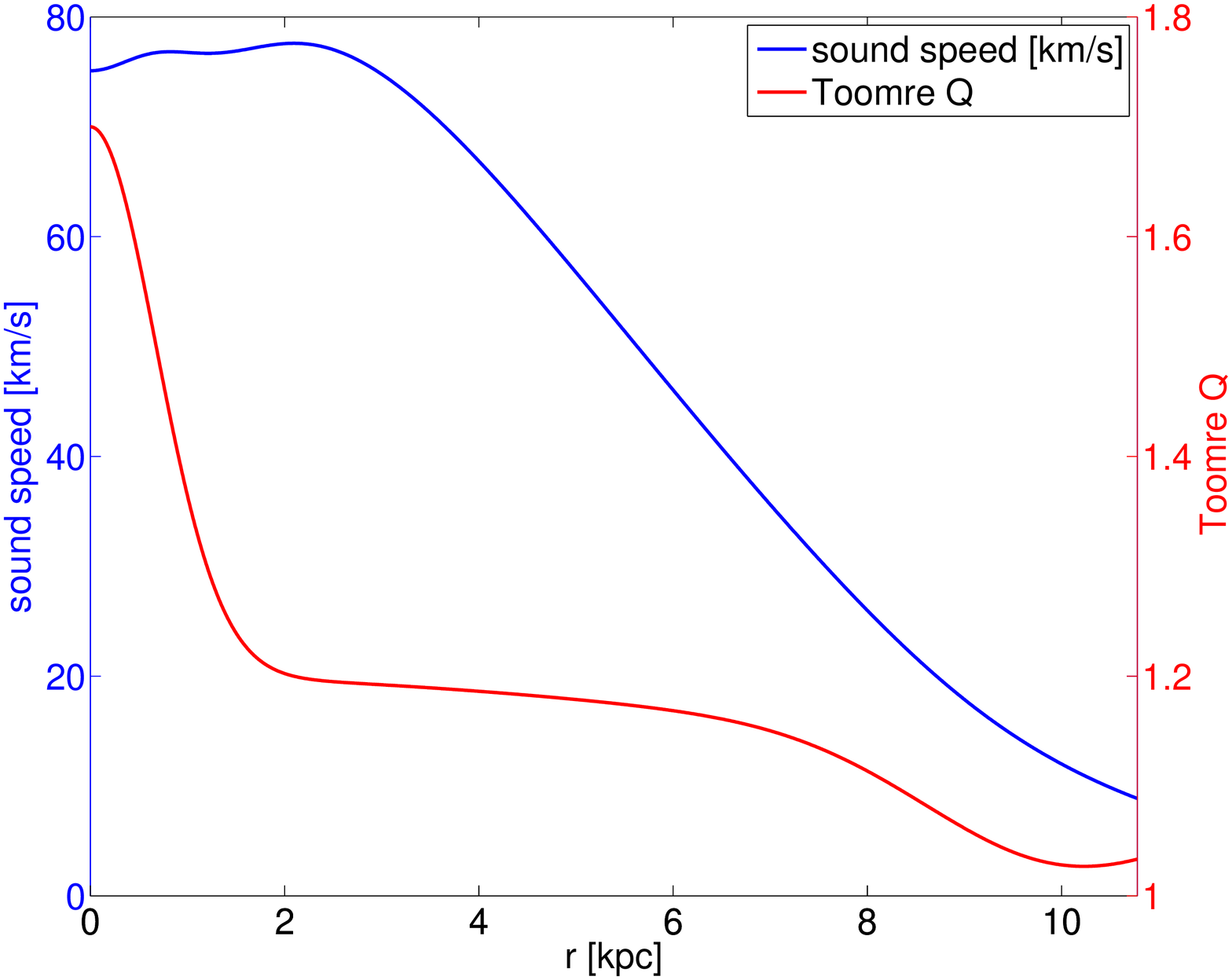}
\caption{The Toomre's $Q$ (right $y$-axis) parameter and the corresponding sound speed (left $y$-axis) 
as functions of galactocenter radius.\label{fig:sound_Q}}
\end{figure}

Aside from the active surface density, $\sigma_0(r)$, and the angular speed, $\Omega(r)$, which can be derived 
from the rotation curve, another constraint associated the basic state is the radial profile of the sound speed, $a(r)$, 
defined within the framework of fluid description. Our knowledge about $a(r)$ is very limited and can only be
inferred from an understanding of the stellar dynamics. The sound speed is related to Toomre's $Q$ through the 
following relation:
\begin{equation} \label{equation:Toomre_Q}
   Q(r) = \frac{a\kappa}{\pi G \sigma_0},
\end{equation}
where $G$ is the gravitational constant and $\kappa$ is the epicyclic frequency. The $Q$ profile provides a 
description of the stability and hotness of the disk. For a zero-thickness disk, $Q<1$ indicates that the disk 
is unstable with respect to axisymmetric perturbations. The disk would quickly evolve and settle down to a 
new basic state, making the description for the original basic state inconsisten. On the other hand, $Q \gg 1$ 
suggests that the disk is dynamically hot and stable. Structures developed in such a disk would not 
survive sufficiently long to account for the frequency of grand-design spiral galaxies. Following these 
dynamical considerations, one can conclude that $Q(r)$ should be greater than but not far from unity. We, 
therefore, adopt the following description for $Q$:
\begin{equation}
   Q=\alpha_0+\Sigma_{i=1}^3 \alpha_i \exp[{-(r-\beta_i)^2/\gamma_i^2}],
\end{equation}
with $\alpha_0=1.0$ and ($\alpha_i,\beta_i,\gamma_i$) summarized in Table~\ref{table:coefficients}. The $Q$ profile 
is composed of a constant term and three Gaussian functions. The coefficients, $\alpha_1,\beta_1,
\gamma_1$, describe a disk which is stable to the axisymmetric perturbation, while the coefficients, $\alpha_2,
\beta_2,\gamma_2$ result in the steep rising $Q$, reflecting the fact that the active disk in the bulge region 
is dynamically hot as compared to the outer disk. The negative contribution from the set, $\alpha_3,\beta_3,\gamma_3$,
is added to adjust the behavior of $k^2_3$ near corotation in order to achieve appropriate growth rates, 
pattern speeds, spiral patterns and spiral amplitudes. This indicates that the behavior of $Q$ near corotation is 
critical to the wave amplification. The resultant $Q$ profile and the corresponding sound speed are shown in 
Fig~\ref{fig:sound_Q}. The sound speed roughly flattens at 76 km~s$^{-1}$ in the bulge region. The 
corresponding values at corotation and the outer 4:1 resonance are 18.0 and 9.0~km~s$^{-1}$, respectively. Due to 
the lack of a direct measurement of the radial velocity dispersion of M81, in the process of modelling, 
the Q profile can be only roughly constrained by the wave phenomenon as seen in the image Fig.~
\ref{fig:M81_residual_map} and guided by the theory of spiral density wave described below. 

\subsection{Modal analysis}

Following \citet{Lau78}, the hydrodynamic equations are formulated in cylindrical coordinates, $(r,\phi, z)$, 
with its origin placed at the galactic center. We consider a density perturbation $\sigma_1$ and all associated 
perturbed quantities are proportional to $\exp[i(\omega t -m\phi+\int^r k dr)]$, with $m$ corresponding to 
the number of spirals and $k$ the radial wave number. Using this representation, negative $k$ traces trailing 
spirals. Treating $\epsilon=a/\kappa r$ as a small parameter, the Poisson equation \citep{Ber79} together with 
the hydrodynamical equations can be systematically expanded to include terms of order $\epsilon^2$. We note that 
this is an excellent approximation for most of the region of our interest except for the very inner part 
($\epsilon>1$ for $r<0.3$~kpc) of the model, where the amplitude of wave is small due to the evanescent 
nature of the wave. Nevertheless, direct comparisons between the results obtained from exact calculations, 
i.e., solving a set of integro-differential equations, and from asymptotic approximation have been shown to be 
in a good agreement \citep{Pan79}. For simplicity, we therefore assume that the governing equation based on 
small $\epsilon^2$ is applicable throughout the region including the galactic center. Consequently, the 
equation we solved for discrete spiral modes is the following differential equation for the enthalpy perturbation 
$h_1=a^2\sigma_1/\sigma_0$ (see the Appendix B in \citet{Lau78} for the details of derivation),
\begin{equation} \label{equation:wave_equation}
\frac{{\rm d}^2u}{dr^2}+k^2_3u =0, \\
\end{equation}
where
\begin{eqnarray}
h_1 = u\left[\frac{\kappa^2(1-\nu^2)}{\sigma_0 r}\right]^{1/2}\exp\left(-i\int^r \frac{\kappa}{aQ} dr \right), \label{equation:transformation} \\
k^2_3 = \hat{k}^2_3+k^2_{\rm co}+k^2_{\rm img}+k^2_{R_1}+k^2_{R_2} \label{equation:k_square}.
\end{eqnarray}
Here, $\nu=(\omega-m\Omega)/\kappa$ is the Doppler-shifted frequency and $u$ is an auxiliary function under 
the transformation Eq.~(\ref{equation:transformation}). The first term on the right of Eq.~(\ref{equation:k_square}), 
which is the major term discussed in \citet{Lau78} is given by
\begin{eqnarray}
\hat{k}^2_3=\left[ \frac{\kappa}{a}\right]^2\left[\frac{1}{Q^2}-1+\nu^2+\frac{1}{4}\mathscr{J}^2Q^2\right], \\
\mathscr{J}= 2m\left(\frac{\pi G \sigma_0}{\kappa^2 r}\right)\left(\frac{1}{s}-\frac{1}{2}\right)^{-1/2}, \label{equation:J} \\
 s = -d\ln \Omega/d \ln r.
\end{eqnarray} 
In addition to $\hat{k}^2_3$, we also consider the term associated with the corotation resonance $k^2_{\rm co}$, 
the ``out of phase" term $k^2_{\rm img}$ and the two remaining terms $k^2_{R_1}$ and $k^2_{R_2}$ defined as follows:

\begin{eqnarray}
 k^2_{\rm co} =-\frac{2\Omega m}{\kappa \nu r^2}\frac{d \ln (\kappa^2/\sigma_0\Omega)}{d \ln r} \label{equation:k_corotation}, \\
 k^2_{\rm img}=-\frac{i\Sigma}{2}\frac{d}{dr} \ln[Q^2(1-\nu^2)], \label{equation:k_img} \\
 k^2_{R_1} = \frac{3}{4r^2}-\left[\frac{r\kappa^2(1-\nu^2)}{\sigma_0}\right]^{1/2}\frac{d^2}{dr^2}\left[\frac{\sigma_0}{r\kappa^2(1-\nu^2))}\right]^{1/2}, \label{equation:k_r1}\\
 k^2_{R_2} =-\frac{4m(\Omega/\kappa)\nu[m\nu r (\Omega'/\kappa)+r\kappa'/\kappa]}{r^2(1-\nu^2)}. \label{equation:k_r2} 
\end{eqnarray}
In Eq.~(\ref{equation:k_corotation}), we have replaced the singular factor $1/\nu$ by the ``plasma dispersion 
relation", $-b\bar{Z}(b\nu)$, when approaching the corotation radius to take into account the smoothing 
effect associated with the epicyclic motions of stars \citep{Mar76a,Mar76b,Lau78,Lau94} where
\begin{equation}
 \bar{Z}(\zeta)=-2i\exp(-\zeta^2)\int^{-i\zeta}_{-\infty}\exp({-t^2})dt,
\end{equation}
and the factor $b$ is defined by:
\begin{equation}
b=\frac{1}{\sqrt{2}r_{\rm co}\epsilon \nu'}.
\end{equation}
Here, $r_{\rm co}$ is corotation radius and $\nu'$ the derivative of $\nu$ with respect to $r$. The connection 
between $1/\nu$ and $-b\bar{Z}(b\nu)$ is rather smooth as in the limit $\epsilon \rightarrow 0$, $-b\bar{Z}(b\nu) 
\rightarrow 1/\nu$. The eigenvalue $\omega=\omega_r+i\omega_i$ of the second-order differential equation 
Eq.~(\ref{equation:wave_equation}) is now determined by imposing the following boundary conditions:
\begin{equation} \label{equation:boundary_inner}
  \frac{du}{dr}=0 {~~~~~~\rm  at~~~~~~} r=0, 
\end{equation}
\begin{equation} \label{equation:boundary_outer}
  \frac{1}{u}\frac{du}{dr}=-ik_3-\frac{1}{2}\frac{d\ln k_3}{dr}  {~~~~~~\rm at ~~~~~~} \nu_r=(\omega_r-m\Omega)/\kappa=0.5. 
\end{equation}
The first condition assumes that the center of the galaxy is a node of the wave, while the second condition 
is a radiation condition, reflecting the absorption of short trailing wave at the outer Lindblad resonance where 
$\nu_r$=1. This choice automatically rules out the possibility of considering any signal from the outer physical 
boundary and implicitly favors trailing waves. Ignoring the possibility of leading waves should not be restrictive
for the purpose of this work, i.e., to provide a dynamical explanation for the grand-design trailing spirals in M81. 
The leading waves are considered to play a minor role in modulating the amplitude of the spirals. Since the 
fluid description of spiral density waves is only appropriate within the principal region defined by $|\nu_r|<1$, 
the choice of the inner and outer boundaries should be sufficiently far from the radii of the Lindblad 
resonances ($|\nu_r|=1$). As a consistency check, there is no ILR associated with our model. Placing the 
outer boundary at the radius of the outer 4:1 resonance is also a reasonable choice for the fluid description.  

In principle, one may obtain solutions by integrating the differential equation Eq.~(\ref{equation:wave_equation}) 
using standard numerical methods if $k^2_3$ is a smooth function over the calculation domain. However, as $r 
\rightarrow 0$, the behavior of the terms $\mathscr{J}$, $k^2_{R_1}$ and $k^2_{R_2}$ depends on the behavior 
of the rotation curve. In general, a singularity would appear at $r=0$, making direct integration numerically 
difficult. To circumvent this problem, either a proper inner boundary condition is imposed at a radius away 
from $r=0$ or the behavior of the rotation curve can remove the singularity. For the later possibility, it 
is straightforward to show that solid body rotation, as modelled in Sec.~2.1, leads to all these terms 
converging to a finite value at $r=0$.

\section{Results}

Given a basic state as described in Sec.~2.1, the governing equation~(\ref{equation:wave_equation}) together 
with the boundary conditions Eqs.~(\ref{equation:boundary_inner})(\ref{equation:boundary_outer}), the eigenvalues, 
$\omega$, and the corresponding eigenfunctions of the system can be found by numerical integration. In 
general, $\omega$ is complex in order to match both the real and imaginary parts at the boundaries. Under the 
assumption of a quasi-stationary spiral structure (QSSS), the pattern speed, $\Omega_p$, is associated 
with the real part through, $\omega_r=m\Omega_p$, while the imaginary part is identified as the temporal 
growth rate of the spiral mode. Unlike the work of \citet{Vis80b}, where the pattern speed, $\Omega_p=18$ km 
s$^{-1}$ kpc$^{-1}$, is constrained by the shape of outer spirals and by some computational considerations, 
within the framework of modal analysis, the pattern speed in this work is determined by the theory and, 
therefore, is consistent with the reaction of the given basic state. In other words, the basic state constructed 
from a real galaxy must be in a configuration such that a proper response and pattern speed is obtained as 
observed.

The solutions of Eq.~(\ref{equation:wave_equation}) can be classified according to the terms of ($m,n$), 
where $m$ is the number of spiral arms and $n$ represents the number of nodes. As the residual map shown in 
Fig.~\ref{fig:M81_residual_map} has nearly bi-symmetric inner and outer spiral arms, we seek eigenvalues 
and eigenfunctions of $m=2$ for the three-component model. For the mode $n=0$, we find $(\omega_r, \omega_i)=
(69.0,-4.8)$, while for $n=1$, $(\omega_r,\omega_i)=(51.0,-4.5)$. These eigenvalues are insensitive to 
the inner boundary condition if Eq.~(\ref{equation:boundary_inner}) is replaced by $u=0$ and to the exact 
location of the outer boundary if it is shifted slightly from the 4:1 resonance. The negative $\omega_i$ 
indicates that these modes are unstable and therefore their amplitudes are exponentially growing with time. 
Direct comparisons between the calculated and the observed data such as the pattern speed, the spiral pattern as well as the spiral amplitude lead us to conclude that the apparent separate inner and outer spiral arms can be explained by a single mode $(m=2,n=1)$. The 
reason that the mode ($m=2,n=1$) dominates over the mode ($m=2,n=0$) can be understood as a result of tidal 
interaction with the companions M82 and NGC 3077 between 200 to 400 Myrs ago \citep{Tho93,Yun92,Yun93}. Since 
the growth rate of these two modes are comparable, the interaction provided a perturbation that favors the 
mode ($m=2,n=1$). 

The pattern speed of this mode is found to be $\Omega_p=25.5$ km s$^{-1}$ kpc$^{-1}$. The corresponding 
radii of corotation and outer 4:1 resonance are located at 9.0~kpc and 10.8~kpc, respectively. Compared with 
the typical pattern speeds quoted in the literature, our value is significantly higher than those that are more 
or less around 18~km~s$^{-1}$~kpc$^{-1}$ \citep{Shu71,Rot75,Vis80a,Vis80b,Got75,Sak87,Ken08}. However, 
our result is in good agreement with those estimated around 25 km s$^{-1}$ kpc$^{-1}$ \citep{Rob75,Elm89,
Low94,Wes98}. The location of the corotation in our model is also consistent with the values around 9~kpc 
quoted in the latter references mentioned above. 

\begin{figure}
\epsscale{1}
\plotone{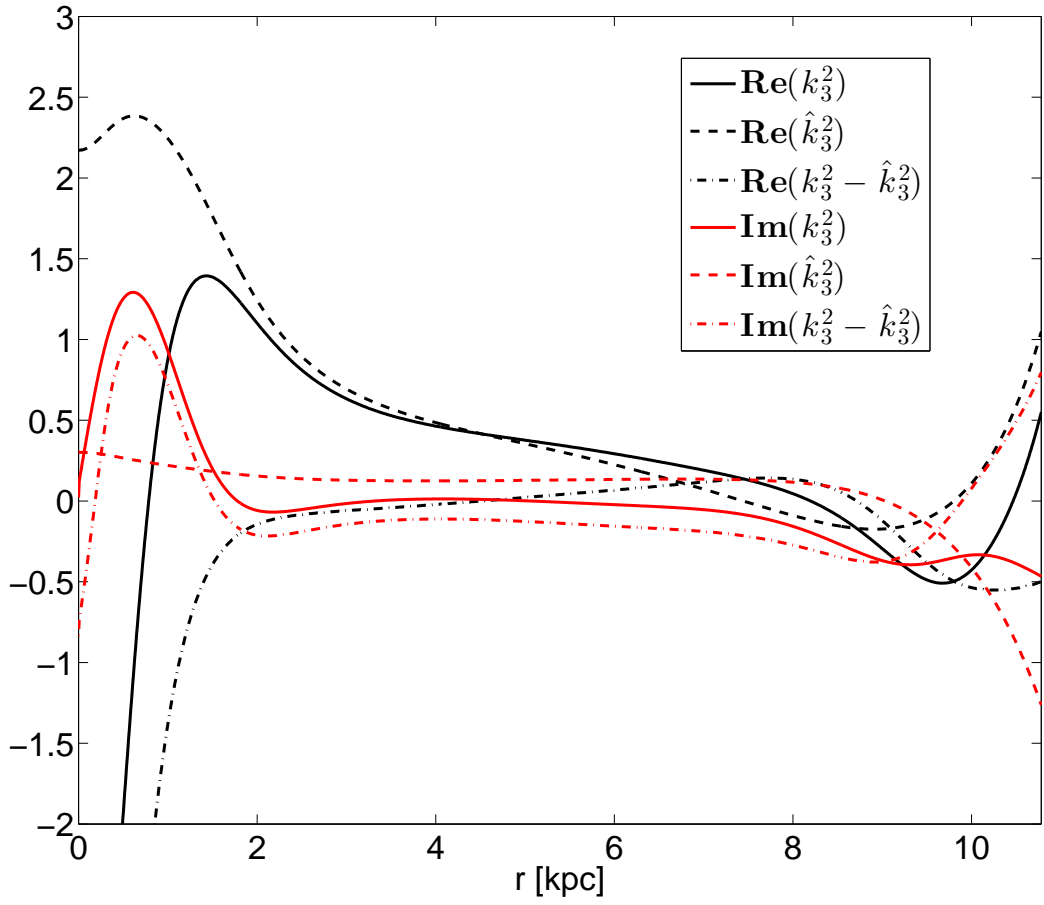}
\caption{The real and imaginary parts of $k^2_3$, $\hat{k}^2_3$ and the sum of remaining terms. \label{fig:k_curve}}
\end{figure}

The radius of the outer 4:1 resonance defines the outer boundary of our calculation. We note the absence of an  
inner Lindblad resonance associated with this pattern speed. Upon substituting this eigenvalue into 
Eq.~(\ref{equation:k_square}), the curves of $k^2_3$, $\hat{k}^2_3$ as well as the sum of remaining terms are 
shown in Fig.~\ref{fig:k_curve}. The real and imaginary parts associated with these quantities are presented 
separately in black and red. From this figure, we first note that in the central region ($r<2$ kpc), both 
$\hat{k}^2_3$ and the remaining terms ($k^2_3-\hat{k}^2_3=k^2_{\rm co}+k^2_{\rm img}+k^2_{R_1} +k^2_{R_2}$) 
contribute to the behavior of the wave. While the term $\mathbf{Re}(\hat{k}^2_3)$ is positive as the 
center is approached, the negative contribution making the $\mathbf{Re}(k^2_3)$ less than zero is 
primarily from $\mathbf{Re}(k^2_{R_2})$ and $\mathbf{Re}(k^2_{\rm co})$. On the other hand, the imaginary part, 
$\mathbf{Im}(k^2_3)$, peaked around $r=0.6$ kpc is mainly from the term $\mathbf{Im}(k^2_{\rm imag}$). As can  
be seen, all the terms on the right of Eq.~(\ref{equation:k_square}) play a role in the analysis. 

From Fig.~\ref{fig:k_curve}, it is important to realize that the imaginary part of $k^2_3$ should not be 
neglected in this model. The condition that one can ignore the imaginary part is that the $Q$-barrier, where 
$\mathbf{Re}(k^2_3)=0$, is sufficiently far away from the region where $\mathbf{Im}(k^2_3)$ is significant. 
This is usually attributable to the fact that the density wave exponentially decays inside the $Q$-barrier, 
where $\mathbf{Re}(k^2_3)<0$, before the term $\mathbf{Im}(k^2_3)$ takes effect.  However, this condition 
does not apply to our model since the `turning point' is located near the peak of $\mathbf{Im}(k^2_3)$. 
As a result, the real and imaginary parts of $u$ are no longer in phase. The concept of $Q$-barrier defined by 
the tuning point in $\mathbf{Re}(k^2_3)$ is not useful in this particular model. Nevertheless, the auxiliary 
function $u$ is forced to decay to zero in our model because of the boundary condition imposed at $r=0$. 

For a given $k^2_3$ and the boundary conditions, $u=u_r+iu_i$ can be determined through numerical integration. 
Here, $(u_r, u_i)$ are the real and imaginary parts of the auxiliary function $u(r)$, respectively.  The 
perturbed enthalpy $h_1$ and the corresponding perturbed surface density, $\sigma_1$, can be calculated 
through the transformation Eq.~(\ref{equation:transformation}) and the relation $\sigma_1=h_1\sigma_0/a^2$. 
In Fig.~\ref{fig:spiral_wave}, we overlap the contours of the calculated density wave for the specific mode 
$(m=2,n=1)$ on the residual image of 3.6~$\mu$m. The good agreement between the observation and the theoretical 
calculation suggests that the inner spirals together with the outer spirals can be explained by the presence 
of a single spiral mode. In particular, they share the same pattern speed 25.5 km s$^{-1}$ kpc$^{-1}$. 
The fact that the seemingly separated inner and outer spirals belong to the same spiral mode is more 
clearly seen in Fig.~\ref{fig:M81_residual_map}. 

\begin{figure}
\epsscale{1}
\plotone{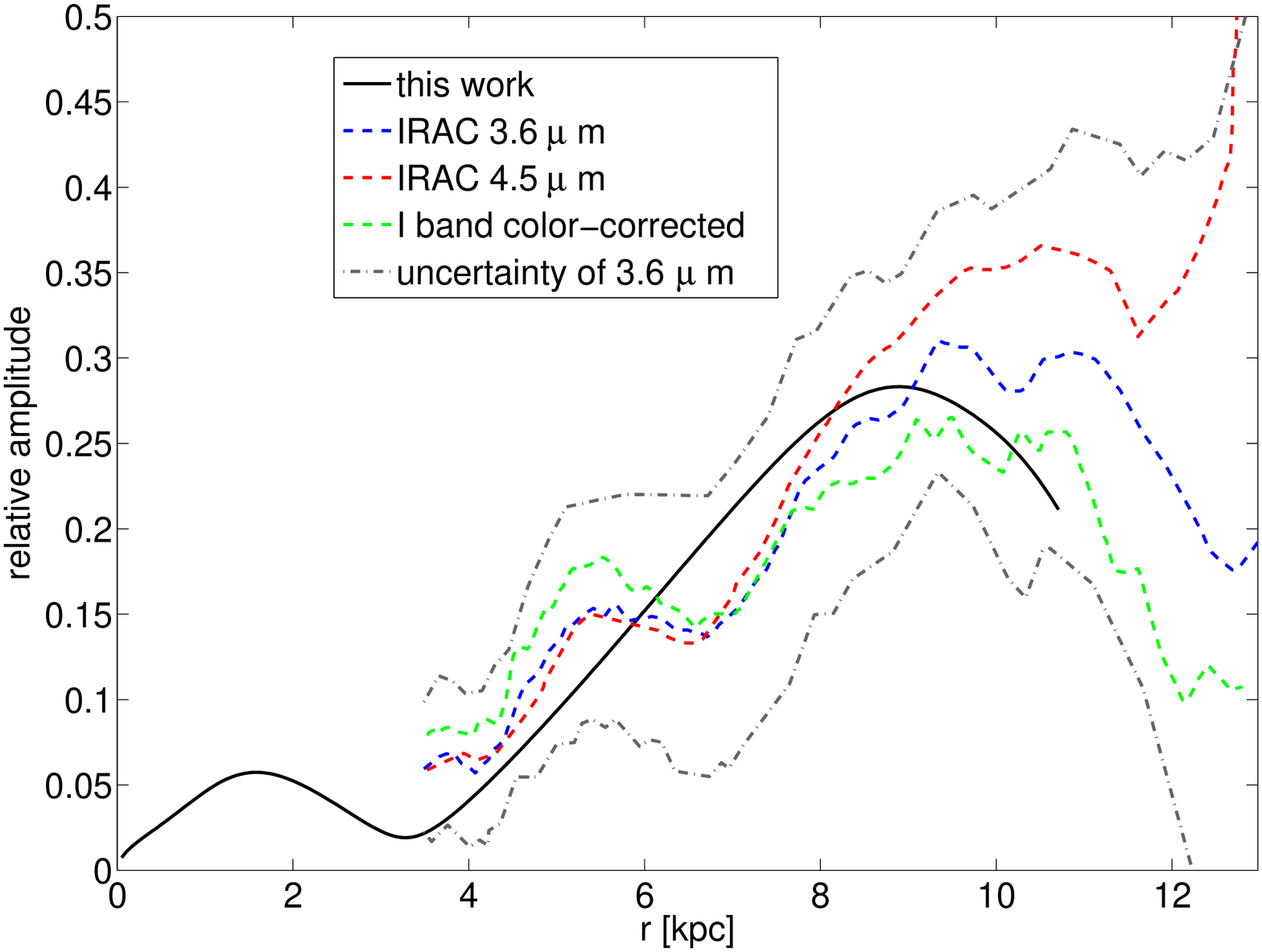}
\caption{Relative amplitude of spiral density wave as a function of radius. The result obtained from linear theory 
of a spiral density wave is scaled to fit the observed data of different wavelengths. Within the range of 
uncertainty (bracketed by the grey dash-dotted curves), the general trend of the calculated waves (solid curve) is in 
good agreement with that of the observed data. The amplitude ripples seen in the observed curves are likely 
due to the interference of trailing waves and leading waves. This interference pattern has been implicitly 
neglected in this work due to the use of a radiation outer boundary condition.\label{fig:amplitude}}
\end{figure}

In Fig.~\ref{fig:amplitude}, the relative amplitude obtained from the mode analysis is rescaled to fit the 
observed IRAC 3.6, 4.5~$\mu$m and {\it I}-band data as shown in \citet{Ken08}. Here, the relative amplitude is 
defined as $\sigma_1/(\sigma_0+\sigma_{\rm b})$, which is consistent with the definition adopted in \citet{Ken08}. 
We note that when calculating the relative amplitude, the projected mass of bulge, $\sigma_{\rm b}$, is taken 
into account since both the disk and the bulge contribute to the axisymmetric part of luminosity. The grey dash-dotted 
lines bracket the uncertainty of the amplitude of IRAC 3.6~$\mu$m. There is no observed information regarding 
the relative amplitude of the inner spirals. The calculated result (solid curve) is shown for the range inside 
the radius of outer 4:1 resonance, which is the domain of calculation. The general trend shows that the observed 
relative amplitude is rising with increasing radius and starts to drop off roughly between 9 to 11~kpc depending 
on the observed wavelength. It is evident that within the range of uncertainty the general trend of the 
calculated amplitude (solid curve) fits the observation reasonably well. The low density contrast in the central 
region ($r<3.5$~kpc) can explain the lack of a detected spiral embedded in the bright bulge. The small 
variations superimposed on the observed amplitude may be interpreted as the interference pattern of leading and 
trailing modes as discussed in \citet{Low94}, which is beyond the scope of this paper.
 
\section{Discussion}

In this work, we interpret the stellar response of a mass model for M81 as an unstable spiral mode in a 
zero-thickness fluid disk. The component that actually participates in the wave propagation is the active stellar 
disk. The bulge and the dark matter halo components are treated as static structures to support the observed 
rotation curve. \citet{Low94} introduce a mass reduction factor $f$ for an exponential disk to incorporate 
the effect of gravity dilution due to the disk thickness. Their active disk density profile is therefore given 
independently of the rotation curve. Despite the fact that the factor $f$ originates from the dynamical 
consideration, we have not introduced this additional adjustable function in modeling the non axisymmetric 
structure for two reasons. First, it is well-known that decomposing the surface light profile into different 
mass components is somewhat ambiguous especially for the bulge region, since the formation process of bulges 
(or pseudo-bulges) and the origin of exponential disks remain unclear. Second, a degeneracy exists in using 
parameters for three-dimensional models in fitting observed two-dimensional data. Thus, we simply adopt  
$f=1$ to minimize the degrees of freedom in the modelling process. In fact, the coefficients ($\alpha_1,
\beta_1,\gamma_1$) that lead to departures of the $Q$-profile from unity implicitly include information 
of the disk thickness. 

Since M81 is not a gas-rich galaxy, we have neglected the limited impact of gas on the stellar dynamics in our 
analysis. However, the dissipative gas may play important roles in maintaining stellar density waves. In addition 
to facilitating the formation of new stars through shock compression \citep{Rob69}, the interstellar medium 
also serves as a coolant that saturates the amplitude of the growing stellar spiral and sustains the conditions 
for the occurrence of grand-design spirals for a relatively long time \citep{Rob72,Ber89,Con86,Pat94}. 
Furthermore, since the density waves in the gas can be transmitted beyond the outer Lindblad resonance, the 
conservation of wave action may lead to the spirals in atomic gas that lie well beyond the optical disk 
\citep{Ber10}. The incompleteness of our model is related to the relatively low sound speed calculated for 
the outer 4:1 resonance, where the surface density of gas may be significant to raise the sound speed to the 
level required for it to be gravitationally stable. Since the equations in our study do not include the 
dissipative nature of gas as well as the interaction between gas and stars, we leave the sound speed as 
assumed. 

The radius of corotation of the spiral pattern in M81 has been determined through various methods as 
summarized in the Section 5.2 of \citet{Ken08}. We do not repeat them here, but comment on the method used by 
\citet{Ken08}, which is based on the phase offset between the stellar spiral and 8-$\mu$m emission. It is known 
from asymptotic analyses as well as from numerical simulations that there exists a angular shift between the peak 
of the gas shock and the potential minimum of the stellar arm \citep{Rob69,Vis80a,Git04}. Here, we 
have implicitly assumed that the dust emission at 8-$\mu$m traces the shock locations. \citet{Git04} 
performed hydrodynamical simulations without self gravity to systematically study the non-linear gas 
response to imposed rigidly rotating stellar spirals. Inside corotation, the gaseous shocks are offset 
in the upstream direction with respect to the potential minimum. Although the actual shift depends on the strength 
of the spirals and the sound speed of the gas, the general trend is that the phase offset increases as 
the corotation radius is approached. Based on the simulation results, they conclude that for two-armed spirals the 
radius of corotation can be determined to within 25 per cent if the angular offset exceeds $\approx \pi/4$. However, the active star formation as observed along spirals suggests that the self gravity of gas is important to the gasdynamics around shocks and therefore may significantly affect the phase shift. The influence of self gravity on the determination of corotation radius based on the phase-offset method will be discussed in a subsequent paper.

`Halo-disk degeneracy' is an outstanding issue when decomposing an observed rotation curve into a luminous disk 
and a dark matter halo \citep{Bershady10}. This is largely due to the uncertainty in the mass-to-light ratio 
and its radial dependence, which depends on the models of star formation history as well as of the initial 
mass function. Furthermore, even for luminous components, the bulge-disk decomposition introduces a further 
uncertainty when modelling the mass distribution of M81. The presence of inner spirals inside the bulge suggests 
the co-existence of disk and bulge in the central region. Despite these difficulties, 
Eq.~(\ref{equation:wave_equation}) associates a proper disk response to a proper surface mass of an `active' 
disk, which might be considered as the lower limit of the disk mass or mass-to-light ratio. As shown in 
Table~\ref{table:mass}, the total luminous mass of our model is about 88\% of the maximum-disk mass, which is 
estimated to be $8.7\times 10^{10}$~M$_{\odot}$ \citep{Ken87}. In order to have the correct stellar disk response, 
it is interesting to note that the mass decomposition of our model is quite similar to that of \citet{Low94} 
and \citet{Ken87}. An additional diagnostic of the mass model may be obtained from the study of the non-linear 
gas response to the observed stellar spiral. Given the observed spiral strength and the distribution of gas, the 
reaction of self-gravitating gas should be compared to the substructures as seen in IRAC-8$\mu$m image. We 
will show in the second paper that only the proper strength of spirals, which is associated with a proper disk 
model, can have an appropriate gas response. 

\hhwang{The absence of inner Lindblad resonance associated with the pattern speeds as reported in this work and inferred from recent observations seems to rule out the spiral-generating mechanism proposed by \citet{Sel12}. There is, however, a possibility that the inner spirals in M81 rotate at a pattern speed different from that of the outer spirals. In this case, the pattern speed of the inner spirals should be lower than 25.5~km~s$^{-1}$
~kpc$^{-1}$. It is also possible that the inner and outer spirals are simply self-perpetuating spiral features as a result of the non-linear development of gravitational wakes \citep{Don13}. As these two sets of 
spirals are virtually separated at the current epoch, it is not immediately clear what these possibilities 
imply for the future evolution of M81. In fact, for this particular case M81, the appearance of inner 
spirals in the nearly uniform rotating part of the disk also poses a challenge for the picture of swing 
amplification in the first place, since the important ingredients for this mechanism such as shear and 
epicyclic motions are no longer coherent in this region. Nevertheless, we emphasize that more than one 
mechanism may operate for spiral galaxies that are in different evolutionary states or environments.}

\hhwang{Although this work is based on several assumptions on the model, three predictions obtained from this particular work can be checked observationally. First, the pattern speed of the inner spirals is the same as the outer spirals. Second, the velocity dispersion as a function of galactocentric distance is constrained by the shape and the strength of spirals. Third, the strength of inner spirals as shown in Fig.~\ref{fig:amplitude} requires observational confirmation.}

\section{Summary}

\begin{figure}
\epsscale{0.5}
\plotone{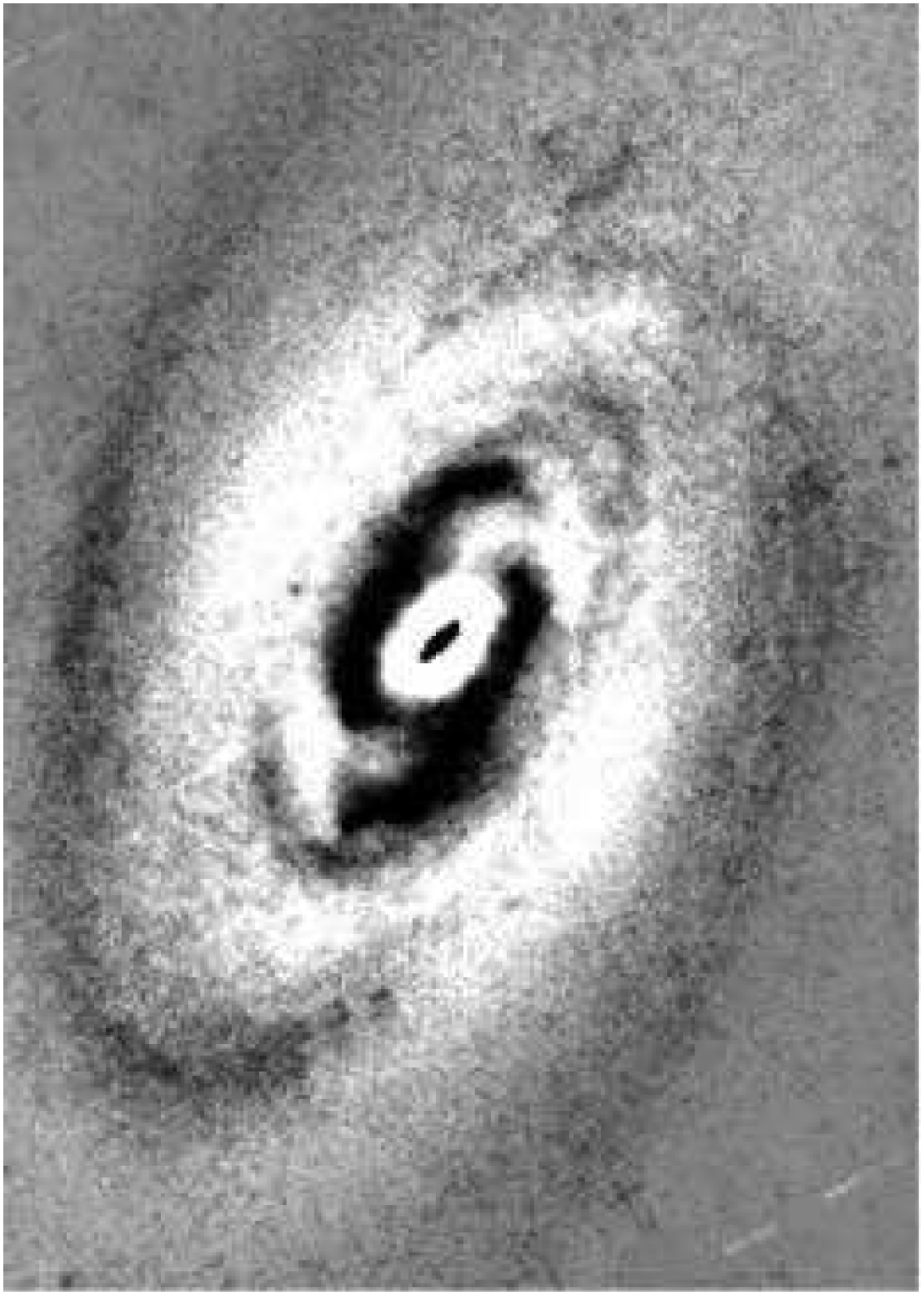}
\caption{The non-axisymmetric residual for the IRAC $3.6~\mu m$ mass map \citep{Ken08}. 
\label{fig:M81_residual_map}}
\end{figure}

The recent discovery of inner stellar spirals lying well inside the bright bulge of M81 cannot be understood 
within the framework of the models proposed by \citet{Vis80b} and \citet{Low94} based on the theory of spiral density 
waves. With little information on the inner spirals in the literature, \citet{Ken08} interpreted that the 
inner stellar arms are probably induced by the density waves in gas. However, from the dynamical point of view, it 
is unlikely that the gas in the bulge region can effectively affect the wave behavior of stars. Hints from the 
image as shown in Fig.~\ref{fig:M81_residual_map} and by the good correlation between dust spirals and stellar 
spirals suggests that the inner spirals together with the grand-design outer spirals may belong to the same spiral 
mode. 

\begin{figure}
\epsscale{0.7}
\plotone{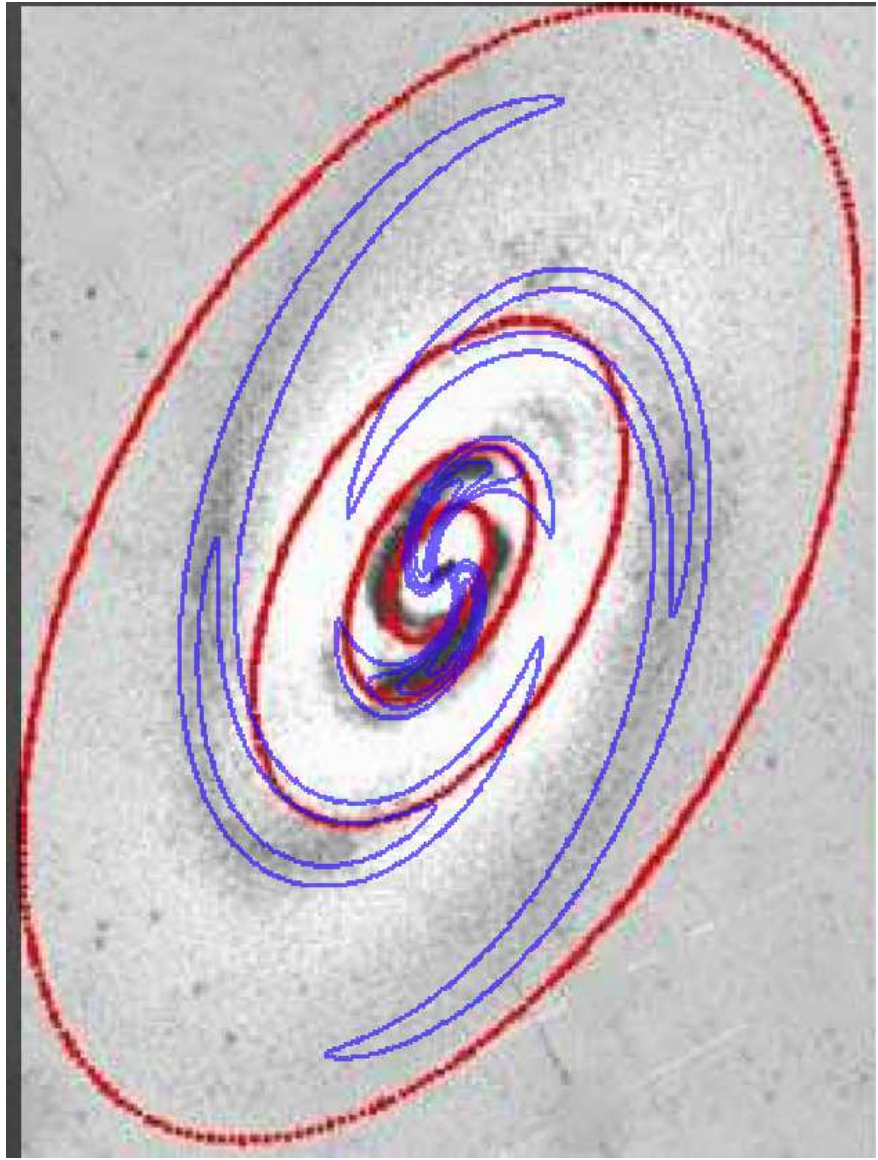}
\caption{The contour map calculated from the mode analysis is overlapped on the residual map of IRAC 3.6~$\mu$m. The red ellipses, with increasing angular size, mark 80, 150, 300 and 675 arcsec, respectively.  \label{fig:spiral_wave}}
\end{figure}

As the first step, a three-component model associated with an active stellar disk, a bulge and a dark matter 
halo is constructed to fit the observed rotation curve and the observed mass distribution. The resultant rotation 
curve is treated as an input function for the mode analysis. Given the boundary conditions, we follow the 
procedure described in \citet{Lau78} and solve for the discrete eigenvalues and the corresponding eigenfunctions. 
In this description, the eigenvalues are associated with pattern speeds and growth rates of the modes and the 
eigenfunctions are related to the spiral patterns. Two dominant unstable modes with comparable growth rates were 
found for $m=2$ in our calculations. The mode ($m=2,n=1$) alone can explain most of the observed characteristics 
of the stellar spirals in M81. Specifically, the pattern speed is determined to be $\Omega_p=25.5$~km~s$^{-1}$
~kpc$^{-1}$ and the corresponding radius of corotation is located at 9~kpc. By directly overlapping the calculated 
density contours on the observation image as shown in Fig.~\ref{fig:spiral_wave}, the apparent unrelated 
inner and outer spirals can be identified as the same spiral mode, rotating rigidly with the same pattern speed. 
Furthermore, the trend of the observed relative amplitude of the spiral can be well reproduced also from the 
mode analysis. This particular mode is likely 'selected' by the three-way interaction with M82 and NGC 3077 
in the not too distant past.

Although the results obtained from this work qualitatively and quantitatively reproduce the observational data 
well, some ingredients associated with a disk galaxy were not taken into account in the mode analysis. For 
example, the effects of the disk thickness and of the interstellar medium are neglected as a first approximation. 
The latter neglect is dynamically important to the outer disk where the surface density of the gas disk is 
comparable to that of the stellar disk. Also the dissipative nature of the gas may be responsible for sustaining and 
regulating the amplitude of stellar spirals. It would be desirable to have measurements of the sound speed (radial velocity dispersion) as a function of radius in as much that it enters into the form of the Toomre Q parameter. The advent of Integral-Field-Unit (IFU) spectroscopy with two-dimensional coverage on the sky has provided the possibility of measuring the stellar velocity ellipsoid over galactic disks \citep{Bershady10}. 
In a subsequent paper, the non-linear gas response to the underlying stellar potential obtained from this 
work will be discussed in detail.

\acknowledgments
\hhwang{The authors would like to acknowledge the support of the Theoretical Institute for Advanced Research in Astrophysics (TIARA) based in Academia Sinica Institute of Astronomy and Astrophysics (ASIAA). We are grateful to G.~Bertin, Y.~Y.~Lau and F.~Shu for the valuable inputs and discussions. The authors thank the referee for comments which helped to improve the clarity and presentation of this paper. Thanks to Mr. Sam Tseng for assistance on the computational facilities and resources (TIARA cluster).}

\bibliographystyle{yahapj}
\bibliography{M81bib}
\clearpage

\begin{table} 
\begin{center}
\caption{Masses of three-component models \label{table:mass}}
\begin{tabular}{clll}
\tableline\tableline 
Mass ($\times 10^{10}$~M$_{\odot}$) & \citet{Low94} & \citet{Ken87} & this work  \\
\tableline
disk   & 5.6 &   6.2 &    6.2 \\
bulge  & 1.3 ($r<12$~kpc)    &   1.0   &    1.5 ($r<12$~kpc)\\
halo  & 3.0 ($r<12$~kpc)&   - &    2.8 ($r<12$~kpc)\\
\tableline
\end{tabular}
\end{center}
\end{table}

\begin{table} 
\begin{center}
\caption{Coefficients associated with the Toomre Q parameter \label{table:coefficients}}
\begin{tabular}{cccc}
\tableline\tableline 
$i$ & 1 & 2 & 3  \\
\tableline
$\alpha_i$ & 0.2 &   0.5 &    -0.1 \\
$\beta_i$ & 0     &   0   &    10.0 \\
$\gamma_i$ & 15.0 &   0.95 &    2.0 \\
\tableline
\end{tabular}
\end{center}
\end{table}
\end{document}